\documentclass[12pt]{iopart}

\usepackage{epsfig}
\usepackage{graphicx}
\usepackage{cite}
\usepackage{subfigure}

\begin{document}
 
\title{Production of resonances in a thermal model: invariant-mass spectra and balance functions }

\author{W.~Florkowski$^{1,2}$, W.~Broniowski$^1$, and P.~Bo\.zek$^1$}

\address{$^1$ The H. Niewodnicza\'nski Institute of Nuclear Physics, 
Polish Academy of Sciences, PL-31342 Krak\'ow, Poland} 
\address{$^2$ Institute of Physics, \'Swi\c{e}tokrzyska Academy,
PL-25406 Kielce, Poland}
 
\begin{abstract}
We present a calculation of the $\pi^+ \pi^-$ invariant-mass correlations and the 
pion balance functions in the single-freeze-out
model. A satisfactory agreement with the data for Au+Au collisions is found. 
\end{abstract}
 

\vspace{-9mm}
\pacs{25.75.-q, 24.85.+p}
\vspace{5mm} 

Thermal models of particle
production in ultra-relativistic heavy-ion collisions turned out to be
very successful in describing hadron yields
\cite{pbmrhic,mich,rafQM02,Bec}, spectra \cite{wbwfspectra,Tor}, the
elliptic-flow coefficients $v_2$ \cite{v2hbt,v2BL}, and the HBT radii
\cite{v2hbt,hbtBL}. In view of this fact it is challenging to study
correlation observables in the thermal approach and compare the results 
to the data. Particular examples of such
observables are the $\pi^+$-$\pi^-$ invariant-mass spectra \cite{starrho}
and the balance functions \cite{bass,jeon}. The latter describe
correlation between opposite-charge particles in the rapidity space
and are closely related to charge fluctuations
\cite{jeonkoch,asakawa,GM}.

In order to analyze the $\pi^+\pi^-$ invariant mass spectra measured in 
Ref.~\cite{starrho} with the like-sign-subtraction technique, we make the assumption that 
all the correlated pion pairs are obtained from the decays of neutral resonances: 
$\rho$, $K_S^0$, $\omega$, $\eta$, $\eta'$, $f_0/\sigma$, and $f_2$. 
The phase-shift formula for the volume density of a $\pi$-$\pi$ resonance with spin degeneracy
$g$ is given by the formula \cite{phsh} 
\begin{equation}
\frac{dn}{dM} = g \int \frac{d^3 p}{(2\pi)^3} \frac{d {\delta_{\pi \pi}(M)}}{\pi dM} 
\left [ \exp \left( \frac{\sqrt{M^2+{p}^2}}{T} \right ) - 1 \right ]^{-1},
\end{equation}
which was used in Ref.~\cite{resonance}, as well as by
Pratt and Bauer \cite{Pratt} in a more recent analysis.
We have performed a calculation in the framework of the single-freeze-out model of 
Ref.~\cite{wbwfspectra}, including the 
flow, experimental kinematic cuts, and decays of higher resonances.
The full results are shown in Ref.~\cite{resonance}. 
In the left panel of Fig.~1 we compare the model predictions to the STAR data. 
The mass of the $\rho$ meson was scaled down by 10\% in accordance to the experimental 
evidence for the dropping mass \cite{starrho,theor}. 
Our results were filtered with the detector efficiency 
correction (we are grateful here to P. Fachini). We can conclude that the model 
does a very good job in reproducing the gross features of the experimental invariant-mass spectra. 
In the right panel of Fig.~1 we show the model predictions for the tranverse momentum spectra 
for several resonances, confronted to data. The 
overall agreement is impressive. Note that all the parameters
of the model (two thermal and two geometric) had been fixed with the help of the spectra of 
pions, kaons, and protons, such that there was no more freedom left in the analysis of resonances.

The balance functions, measured by the STAR Collaboration
\cite{STARbal,phdmsu}, are defined by the formula
\begin{equation}
B(\delta,Y) = {1\over 2} 
\left\{
{\langle N_{+-}(\delta) \rangle - \langle N_{++}(\delta) \rangle \over
\langle N_+ \rangle} 
+
{\langle N_{-+}(\delta) \rangle - \langle N_{--}(\delta) \rangle \over
\langle N_- \rangle} \right\},
\label{def}
\end{equation} 
where $N_{+-}(\delta)$ denotes the number of the opposite-charge pairs such
that both members of the pair fall into the rapidity window $Y$ with
relative rapidity $|y_2-y_1|=\delta$, $N_{+}$ is the number
of positive particles in the interval $Y$, {\em etc.}
The measurement \cite{STARbal} showed that the widths of the balance
functions are smaller than expected from models discussed in
Ref.~\cite{bass}.  This problem was discussed by Bialas in
Ref. \cite{Bialas} in the framework of the quark coalescence model \cite{coal}.
Here we present the results of the calculation of
the $\pi^+ \pi^-$ balance function \cite{ourbal} based on the
model or Ref.~\cite{wbwfspectra}.  In a more recent
paper Pratt {\em et al.} \cite{many} showed that the measured
balance function may also be satisfactorily reproduced in the blast-wave
model.

\begin{figure}[tb]
\begin{center}
\subfigure{\includegraphics[angle=0,width=0.48\textwidth]{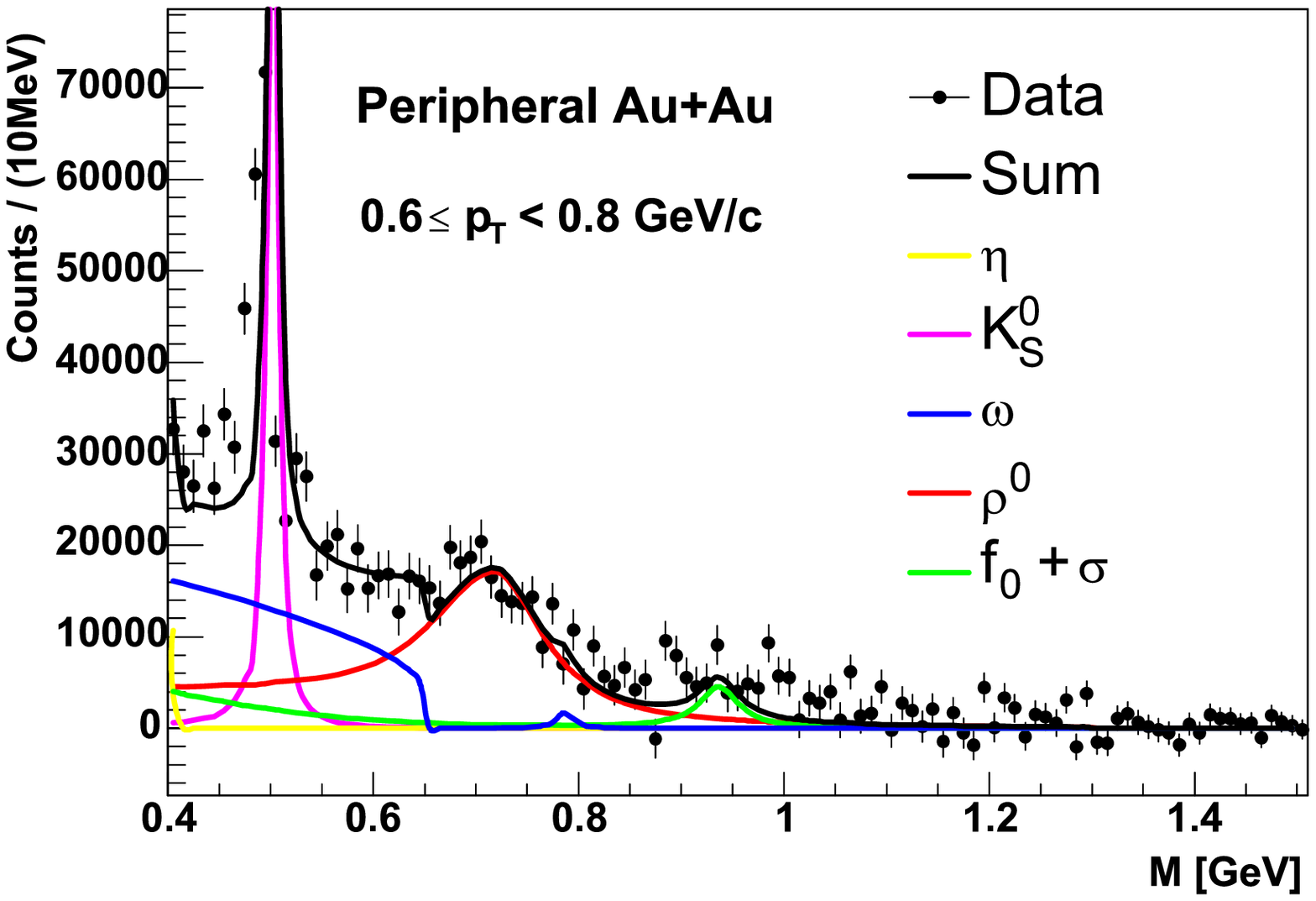}}
\subfigure{\includegraphics[angle=0,width=0.48\textwidth]{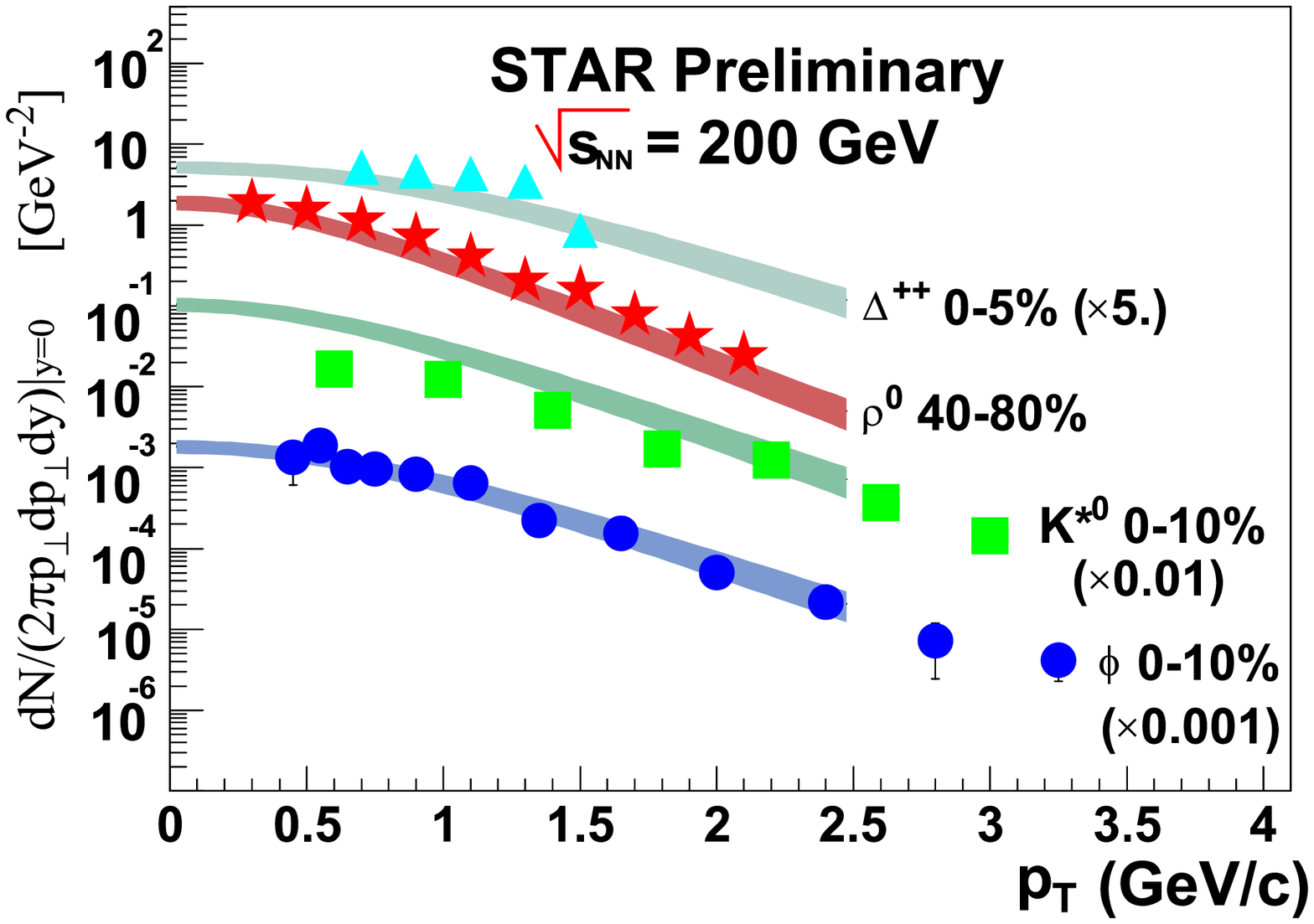}}\\
\end{center}
\vspace{-9mm}
\caption{Left: single-freeze-out model vs. the data of Ref.~\cite{starrho} for the $\pi^+ \pi^-$ 
invariant-mass correlations. The model calculation includes the 
decays of higher resonances, the flow, the kinematic cuts, and the detector efficiency. Right: 
parameter-free 
model predictions and data for the transverse-momentum spectra of several resonances.}
\vspace{-5mm}
\end{figure}

In our approach the $\pi^+ \pi^-$ balance function has two
contributions related to two different mechanisms of the creation of
an opposite-charge pair. The first one {\it (resonance contribution)}
comes from decays of neutral hadronic resonances, whereas
the second one {\it (non-resonance contribution)} is related to other
possible correlations. We assume that the
second mechanism forces the two opposite-charge pions to be produced
at the same space-time point and with thermal velocities. 
The first contribution includes {\it
neutral} resonances which have a $\pi^+ \pi^-$ pair in the final state
(we explicitly include $K_S$, $\eta$, $\eta^\prime$, $\rho^0$,
$\omega$, $\sigma$, and $f_0$); here the
correlations are completely determined by the
kinematics. The details of the model and other technical remarks are given in
Ref. \cite{ourbal}.

\begin{figure}[tb]
\begin{center}
\hspace{0.5cm} \epsfxsize=13.5cm \epsfbox{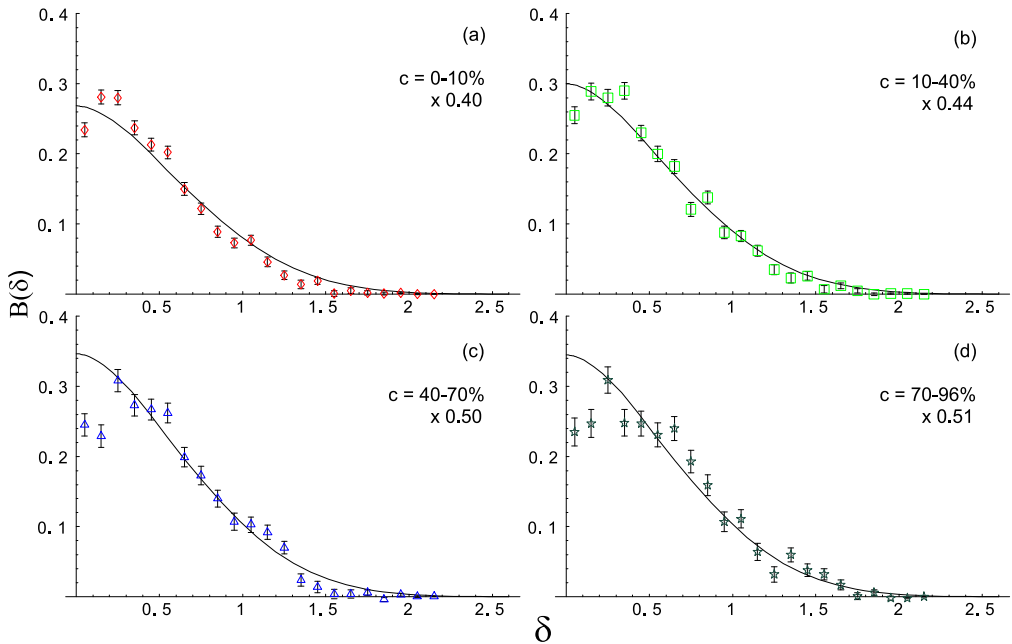}
\end{center}
\vspace{-7mm}
\caption{Balance functions for the pions in the thermal model
calculated for four different centrality classes and compared to the
experimental data of Ref.~\cite{STARbal}. The normalization of the
model curves was adjusted in each case and is listed near the plot
labels.}
\label{thdat}
\vspace{-0.5cm}
\end{figure}

According to the above discussion, the $\pi^+ \pi^-$
balance function can be constructed as a sum of the two terms,
\begin{equation}
B(\delta) = B_{\rm R}(\delta) + B_{\rm NR}(\delta).
\label{balsum}
\end{equation}
The functions $B_{\rm R}(\delta)$ and $B_{\rm NR}(\delta)$ resulting
from our model calculation are presented in Ref.~\cite{ourbal}. The
value of the temperature used in the calculation was $T = 165$~MeV,
and the expansion parameters of the model had been fitted to the spectra
of hadrons \cite{wbwfspectra}. This procedure yields the average transverse flow of
0.5$c$. One can observe that the widths of the two contributions are
similar. The calculated total width, $\langle \delta \rangle = 0.66 $,
turns out to be somewhat larger than the experimental value for the
most central collisions. The STAR result for the most central
events ($c=0-10\%$) is $\langle \delta \rangle = 0.594\pm 0.019$, for
the mid-central ($c=10-40\%$) $\langle \delta \rangle = 0.622\pm
0.020$, for the mid-peripheric ($c=40-70\%$) $\langle \delta \rangle =
0.633\pm 0.024$, and for the peripheric ($c=70-96\%$) $\langle \delta
\rangle = 0.664\pm.029$. Such dependence of the width of the balance
function on centrality cannot be reproduced in our model via changing
the transverse flow within the limits consistent with the single-particle
spectra.

In Fig. 2 our results are compared to the experiment. The normalization of
the model curves was adjusted in each case, since we were not able to
take into account, in a more sophisticated way, the effect of the limited
detector efficiency and acceptance. On the other hand, the kinematic
cuts in pseudorapidity and transverse momentum were included 
\cite{ourbal}. The shapes of the model balance functions agree well
with the data except for the most central case where the theoretical
width is slightly larger. The dips of the experimental
balance functions at very small values of $\delta$ are caused by the
HBT correlations, not included in our
approach.  We also note that the effects of the detector efficiency
may influence the width of the balance function \cite{many}.

In conclusion, we state that the single-freeze-out model gives a
satisfactory description of the pion invariant-mass correlations and 
of the balance functions. The results for the transverse-momentum
spectra of resonances are in remarkable overall agreement with the experiment.

We are grateful to Patricia Fachini for numerous helpful discussions and for preparing the figures.
This research has been supported in part by the Polish State 
Committee for Scientific Research grant 2 P03B 059 25.

\end{document}